\documentclass[aps,prb,groupedaddress,showpacs]{revtex4}  
\usepackage{graphicx}  

\def\be{\begin{equation}}  
\def\ee{\end{equation}}  
\def\ba{\begin{eqnarray}}  
\def\ea{\end{eqnarray}}  
\def\bc{\begin{center}}  
\def\ec{\end{center}}  
\def\p{\partial}  
  
\def\Sp{{\rm Sp}}

\begin{document}  

\title{Theory of the giant plasmon enhanced second harmonic generation in graphene and semiconductor two-dimensional electron systems}  

\author{S. A. Mikhailov}  
\email[Electronic mail: ]{sergey.mikhailov@physik.uni-augsburg.de}  

\affiliation{Institute of Physics, University of Augsburg, D-86135 Augsburg, Germany}  

\date{\today}  

\begin{abstract}  
An analytical theory of the nonlinear electromagnetic response of a two-dimensional (2D) electron system in the second order in the electric field amplitude is developed. The second-order polarizability and the intensity of the second harmonic signal are calculated within the self-consistent-field approach both for semiconductor 2D electron systems and for graphene. The second harmonic generation in graphene is shown to be about two orders of magnitude stronger than in GaAs quantum wells at typical experimental parameters. Under the conditions of the 2D plasmon resonance the second harmonic radiation intensity is further increased by several orders of magnitude.  
\end{abstract}  

%\keywords{}  
% 78.67.Wj 	Optical properties of graphene 
%73.20.Mf 	Collective excitations (including excitons, polarons, plasmons and other charge-density excitations)
%42.65.Ky 	Frequency conversion; harmonic generation, including higher-order harmonic generation

\maketitle  

\section{Introduction}

Graphene is a recently discovered \cite{Novoselov04} purely two-dimensional (2D) material consisting of a monolayer of $sp^2$-bonded carbon atoms arranged in a hexagonal lattice. Electrons and holes in graphene are massless Dirac fermions and this leads to a variety of interesting and unusual electrical and optical properties of this material\cite{Novoselov05,Zhang05,Geim07a,Geim09,Neto09,Bonaccorso10}. It promises a lot of applications in electronics, optics and other areas\cite{Geim07a,Geim09,Neto09,Bonaccorso10}.

It has been predicted \cite{Mikhailov07e} that the unusual linear energy dispersion of the charge carriers should lead to a {\em strongly nonlinear} electromagnetic response of graphene: the irradiation by electromagnetic waves should stimulate the emission of higher frequency harmonics from graphene. The theory of the nonlinear electromagnetic response of graphene has been further developed in Refs. \cite{Mikhailov08a,Mikhailov08b,Lopez08,Mikhailov09a,Wright09,Ishikawa10,Mikhailov10,Mikhailov11a}. Experimentally the higher harmonics generation and frequency mixing effects have been observed in Refs. \cite{Wang09,Dean09,Dragoman10,Hendry10}. 

In theoretical papers \cite{Mikhailov08a,Mikhailov08b,Lopez08,Mikhailov09a,Wright09,Ishikawa10,Mikhailov10,Mikhailov11a} only the normal incidence of radiation on a uniform graphene layer has been studied. The experimental demonstration \cite{Hendry10} of the third-order emission of radiation at the frequency $2\omega_1 - \omega_2$ at the bichromatic irradiation by the frequencies $\omega_1$ and $\omega_2$ has confirmed that graphene manifests the nonlinear properties and that its third-order effective nonlinear susceptibility is much higher than in a number of other materials \cite{Hendry10,Mikhailov10,Mikhailov11a}. However, the intensity of the emitted signal $I_{2\omega_1\pm\omega_2}\propto I_{\omega_1}^2I_{\omega_2}$ is proportional to the third power of the intensities of the incident waves \cite{Hendry10} therefore to observe the third-order nonlinear effects one needs quite powerful sources of radiation.

Substantially stronger nonlinear effects could be expected in the second order in the radiation electric field. The second-order effects, e.g. the second harmonic generation, are proportional to the second power of the incident wave intensities, $I_{2\omega}\propto I_{\omega}^2$. However graphene is a centrosymmetric material, therefore at the normal incidence of radiation the second-order effects are forbidden by symmetry. 

The symmetry arguments do not hinder the observation of the second-order effects at the oblique incidence of radiation on the 2D electron layer. If the incident wave has the wavevector component ${\bf q}$ parallel to the plane of the 2D layer, one could observe much stronger second-harmonic radiation  as compared to the third order effects \cite{Hendry10}. Moreover, at the oblique incidence of radiation one can resonantly excite the 2D plasma waves in the system (e.g. in the attenuated total reflection geometry) which would lead to the resonant enhancement of the higher harmonics \cite{Simon74}. 

In this paper we theoretically study the second-order nonlinear electromagnetic response of two-dimensional electron systems, including both graphene and conventional 2D structures with the parabolic electron energy dispersion. We calculate the second-order polarizability $\alpha^{(2)}$ of 2D electrons (Section \ref{sec:resp}) and show that in graphene it is {\em at least one order of magnitude larger} than in typical semiconductor structures (e.g. in GaAs/AlGaAs quantum wells). Then we calculate the self-consistent response of the system to an external radiation taking into account the 2D plasmon excitation (Section \ref{sec:self}) and show that the intensity of the second harmonic signal can be further increased {\em by several orders of magnitude}. In Section \ref{sec:summ} we summarize our results.

\section{Response equations\label{sec:resp}} 

\subsection{General solution of the quantum kinetic equation} 

We consider a 2D electron system with the Hamiltonian $\hat H_0$, where 
\be 
\hat H_0|\lambda\rangle=E_\lambda|\lambda\rangle,
\ee
$\lambda$ is a set of quantum numbers, $E_\lambda$ and $|\lambda\rangle$ are the eigenenergies and eigenfunctions of the system. Under the action of the electric field described by the scalar potential
\be 
\phi({\bf r},t)=
\phi_{{\bf q}\omega} e^{i({\bf q}\cdot {\bf r}-\omega t)} + c.c. \label{pot}
\ee
the state of the system is perturbed (here c.c. means the complex conjugate). 
Its response to the potential (\ref{pot}) is determined by the Liouville equation 
\be 
i\hbar \frac{\p \hat{\rho}}{\p t}=[\hat H,\hat \rho] =[\hat H_0+\hat H_1,\hat \rho] 
\ee 
where $\hat \rho$ is the density matrix and $\hat H_1=-e\phi({\bf r}, t)$. Our goal is to calculate the charge density fluctuations 
\be
\rho({\bf r},t)=\rho_{{\bf q}\omega} e^{i({\bf q}\cdot {\bf r}-\omega t)} + \rho_{2{\bf q},2\omega} e^{2i({\bf q}\cdot {\bf r}-\omega t)} + c.c.
\ee
in the first and second orders in the potential amplitudes $\phi_{{\bf q}\omega}$ and the corresponding first- and second-order polarizabilities $\alpha^{(1)}_{{\bf q}\omega;{\bf q}\omega}$ and $ \alpha^{(2)}_{2{\bf q},2\omega;{\bf q}\omega,{\bf q}\omega}$, defined as 
\be
\rho_{{\bf q}\omega}=\alpha^{(1)}_{{\bf q}\omega;{\bf q}\omega}\phi_{{\bf q}\omega},\label{a1-def}
\ee
\be
\rho_{2{\bf q},2\omega} = \alpha^{(2)}_{2{\bf q},2\omega;{\bf q}\omega,{\bf q}\omega}\phi_{{\bf q}\omega}\phi_{{\bf q}\omega}.\label{a2-def}
\ee

In the absence of the perturbation $\hat H_1$ the density matrix $\hat \rho_0$ satisfies the equation 
\be 
\hat \rho_0|\lambda\rangle=f_\lambda|\lambda\rangle,
\ee
where $f_\lambda=f(E_\lambda)$ is the Fermi distribution function. 
Expanding $\hat \rho$ 
in powers of the electric potential, $\hat \rho=\hat \rho_0+\hat \rho_1+\hat \rho_2+\dots$, and calculating the charge density fluctuations $- e\Sp \left[\delta({\bf r-r}_0)\left(\hat\rho_1+\hat\rho_2+\dots\right)\right]$ 
we get 
\be 
\alpha^{(1)}_{{\bf q}\omega;{\bf q}\omega}=\frac {e^2}S
\sum_{\lambda\lambda'} 
\frac{f_{\lambda'}-f_{\lambda}} {E_{\lambda'}-E_{\lambda}+\hbar\omega+i0}
\langle \lambda'|e^{-i{\bf q}\cdot {\bf r}}|\lambda\rangle
\langle\lambda|e^{i{\bf q}\cdot {\bf r}} |\lambda'\rangle ,
\label{a1}
\ee 
\ba
\alpha^{(2)}_{2{\bf q}2\omega;{\bf q}\omega,{\bf q}\omega}
&=&
-\frac {e^3}S\sum_{\lambda\lambda'} 
\frac{\langle \lambda'|e^{-i2{\bf q}\cdot {\bf r}}|\lambda\rangle}{E_{\lambda'}-E_{\lambda}+2\hbar\omega +2i0}
\sum_{\lambda''} 
\langle\lambda|e^{i{\bf q}\cdot {\bf r}}|\lambda''\rangle \langle\lambda''|e^{i{\bf q}\cdot {\bf r}} |\lambda'\rangle
\nonumber \\ &\times&
\Bigg( 
\frac{f_{\lambda'}-f_{\lambda''}} {E_{\lambda'}-E_{\lambda''}+\hbar\omega+i0}
- 
\frac{f_{\lambda''}-f_{\lambda}} {E_{\lambda''}-E_{\lambda}+\hbar\omega+i0}
\Bigg),
\label{a2}
\ea
where $S$ is the sample area. The first-order polarizability (\ref{a1}) is proportional to the polarization operator $\Pi({\bf q},\omega)$ (see \cite{Stern67}), $\alpha^{(1)}_{{\bf q}\omega;{\bf q}\omega}=-e^2\Pi({\bf q},\omega)$. For the conventional 2D electron gas (with the parabolic electron energy dispersion) the linear polarizability (\ref{a1}) has been calculated in Ref. \cite{Stern67}, for 2D electrons in graphene (with the linear energy dispersion) it has been done in Refs. \cite{Hwang07,Wunsch06}. 

We will apply the general formulas (\ref{a1}), (\ref{a2}) to the conventional 2D electron systems in semiconductor heterostructures and to graphene. In the former case the spectrum of 2D electrons is parabolic, in the latter case it is linear. In both cases the set of quantum numbers $|\lambda\rangle=|l{\bf k}\sigma\rangle$ consists of the subband index $l$, the wavevector ${\bf k}$ and the spin $\sigma$. To specify the general expressions (\ref{a1}), (\ref{a2}) we consider the long-wavelength limit, which is quantitatively described by the conditions  
\be 
q\ll \max\{k_F,k_T\},\ \  q\ll\omega/\max\{v_F,v_T\} 
\label{smallqcond-par}
\ee 
in semiconductor 2D electron systems and the conditions 
\be 
q\ll \max\{k_F,k_T\},\ \  q\ll\omega/v_F 
\label{smallqcond-lin}
\ee 
in graphene. Here $k_F$ and $v_F$ are the Fermi wave-vector and Fermi velocity respectively, $k_T$ and $v_T$ are the thermal wave-vector and velocity respectively. In the gas with the parabolic dispersion $k_F=\sqrt{2mE_F}/\hbar$, $k_T=\sqrt{2mT}/\hbar$, $v_{F,T}=\hbar k_{F,T}/m$, where $m$ is the effective electron mass, $T$ is the temperature and the Fermi energy $E_F$ is counted from the bottom of the parabolic band. In the 2D gas with the linear energy dispersion (in graphene) $k_F=\sqrt{|\mu|}/\hbar v_F$, $k_T=\sqrt{T}/\hbar v_F$, where $\mu$ is the chemical potential counted from the Dirac point ($\mu$ can be positive and negative) and $v_F$ is the Fermi velocity. At typical experimental parameters the conditions (\ref{smallqcond-par}) and (\ref{smallqcond-lin}) restrict the wavevector $q$ by the values $\sim 10^6$ cm$^{-1}$ which is sufficient for the description of most experiments. 

Under the conditions (\ref{smallqcond-par}) -- (\ref{smallqcond-lin}) the general expressions (\ref{a1}) and (\ref{a2}) can be substantially simplified and we get the first and second order polarizabilities of the 2D electron gas in the form
\be 
\alpha^{(1)}_{{\bf q}\omega;{\bf q}\omega}\approx
\frac {e^2 q_\alpha q_\beta}{(\hbar\omega)^2 } \frac {g_s}S
\sum_{l{\bf k}} \left(-\frac{\p f_{l{\bf k}}}{\p k_\alpha}\right)
\frac {\p E_{l{\bf k}}}{\p k_\beta},\label{a1-expand}
\ee 
\be
\alpha^{(2)}_{2{\bf q}2\omega;{\bf q}\omega,{\bf q}\omega}
\approx
\frac {3e^3 q_\alpha q_\beta q_\gamma q_\delta }{2(\hbar\omega)^4 }
\frac{g_s}S\sum_{l{\bf k }} 
\frac{\p f_{l{\bf k}}}{\p k_\alpha}
\frac {\p E_{l{\bf k}}}{\p k_\beta}
\frac{\p^2 E_{l{\bf k}}}{\p k_\gamma \p k_\delta},
\label{a2-expand}
\ee
where $g_s=2$ is the spin degeneracy.

\subsection{2D electron gas with the parabolic energy dispersion}

Let us apply now the obtained formulas (\ref{a1-expand}) and (\ref{a2-expand}) to the conventional 2D electron gas with the parabolic energy dispersion. In this case there is only one energy subband ($l=1$), 
\be
E_{l{\bf k}}\equiv E_{\bf k}=\frac{\hbar^2 k^2}{2m} \label{energyparab}
\ee
and the wave functions are plane waves. Substituting  (\ref{energyparab}) in Eqs. (\ref{a1-expand}) and (\ref{a2-expand}) we get 
\be
\alpha^{(1)}_{{\bf q}\omega;{\bf q}\omega}\approx 
\frac {n_se^2q^2}{m\omega^2}
\label{alpha1par-final}
\ee 
and 
\be
\alpha^{(2)}_{2{\bf q}2\omega;{\bf q}\omega,{\bf q}\omega}
\approx 
-\frac {3n_se^3q^4 }{2m^2\omega^4}.
\label{alpha2par-final}
\ee
In semiconductor 2D electron systems both $\alpha^{(1)}$ and $\alpha^{(2)}$ are proportional to the 2D electron gas density $n_s$. The result (\ref{alpha1par-final}) has been obtained in Ref. \cite{Stern67}.

\subsection{2D electron gas with the linear energy dispersion (graphene) }

In graphene the spectrum of electrons can be found in the tight-binding approximation \cite{Wallace47}. It consists of two energy bands, the wave functions are described by Bloch functions and the energy of the electrons is
\be
E_{l{\bf k}}=(-1)^lt |{\cal S}_{\bf k}|, \ \ l=1,2,
\label{energy}
\ee
where $t$ is the transfer integral, ${\cal S}_{\bf k}$ is a complex function defined as 
\be 
{\cal S}_{\bf k}=1 
+e^{i{\bf k}\cdot{\bf a}_1} 
+e^{i{\bf k}\cdot{\bf a}_2}=1+2\cos(k_xa/2)e^{i\sqrt{3}k_ya/2},\label{calS}
\ee
${\bf a}_1=a(1/2,\sqrt{3}/2)$, ${\bf a}_2=a(-1/2,\sqrt{3}/2)$ are the basis vectors of the graphene hexagonal lattice and $a=|{\bf a}_1|=|{\bf a}_2|$ is the lattice constant. In graphene $t\approx 3$ eV and $a=2.46$ \AA.

The formulas (\ref{a1})--(\ref{a2}) and (\ref{a1-expand})--(\ref{a2-expand}) are valid for the full graphene energy dispersion, i.e. the integration in these formulas is performed over the whole Brillouin zone. Under the real experimental conditions, when the chemical potential $\mu$ satisfies the condition $|\mu|\ll t$, the main contribution to the integrals (\ref{a1-expand}) and (\ref{a2-expand}) is given by the vicinity of two Dirac points ${\bf K}_1=(2\pi/a)(1/3,1/\sqrt{3})$, ${\bf K}_2=(2\pi/a)(2/3,0)$, where the function ${\cal S}_{\bf k}$ vanishes, ${\cal S}_{{\bf K}_1}={\cal S}_{{\bf K}_2}=0$, and the energy (\ref{energy}) can be approximated by linear functions 
\be 
E_{l{\bf k}}=(-1)^l \hbar v_F|\tilde {\bf k}|=(-1)^l \hbar v_F|{\bf k-K}_v|,\ \ v=1,2,\label{energy-lin}
\ee 
where $v$ is the valley index and $\tilde {\bf k}={\bf k-K}_v$ is the electron wavevector counted from the corresponding Dirac points. 
The velocity parameter $v_F$ here (the Fermi velocity) is related to the transfer integral and the lattice constant, $v_F=\sqrt{3}ta/2\hbar$; in graphene $v_F\approx 10^8$ cm/s. Omitting below the tilde over the wavevector ${\bf k}$ and calculating the derivatives 
\be
\frac {\p E_{l{\bf k}}}{\p k_\alpha}=(-1)^l\hbar v_F\frac {k_\alpha}k , \ \ \ 
\frac {\p^2 E_{l{\bf k}}}{\p k_\alpha\p k_\beta}=(-1)^l\hbar v_F\frac{k^2\delta_{\alpha\beta}-k_\alpha k_\beta}{k^3}\label{deriv}
\ee
we get the first order polarizability in the form
\be 
\alpha^{(1)}_{{\bf q}\omega;{\bf q}\omega}=
\frac {e^2g_sg_vq^2 T}{4\pi \hbar^2\omega^2 }
\left[ F\left(\frac {|\mu|}{2T}\right) 
+F\left(-\frac {|\mu|}{2T}\right) 
\right],\label{gr-a1}
\ee
where $g_v=2$ is the valley degeneracy and 
\be 
F(a)=\int_0^\infty \frac{xdx}{\cosh^2(x-a)}\approx 
\left\{
\begin{array}{ll}
2a, & a\gg 1 \\
\ln 2 , & a=0 \\
\end{array}
\right. .
\ee
If the temperature is low as compared to the chemical potential, $T\ll|\mu|$ we get from (\ref{gr-a1}) the result obtained in Refs. \cite{Hwang07,Wunsch06},
\be
\alpha^{(1)}_{{\bf q}\omega;{\bf q}\omega}=
\frac {e^2g_sg_vq^2 |\mu| }{4\pi \hbar^2\omega^2 },\ \ T\ll |\mu|.
\ee
In the opposite case $|\mu|\ll T$ one has
\be
\alpha^{(1)}_{{\bf q}\omega;{\bf q}\omega}=
\frac {e^2g_sg_vq^2T }{4\pi \hbar^2\omega^2 }
\left( 2\ln 2+\left(\frac \mu{2T}\right)^2 \right)
, \ \ |\mu|\ll T. 
\ee

Now consider the second order polarizability of graphene. Using Eqs. (\ref{a2-expand}) and (\ref{deriv}) we get
\be
\alpha^{(2)}_{2{\bf q}2\omega;{\bf q}\omega,{\bf q}\omega}
\approx
-\frac {3e^3g_sg_v q^4 v_F^2}{32\pi \hbar^2\omega^4 }
\tanh \frac{\mu}{2T}.\label{gr-a2}
\ee
The second-order graphene polarizability $\alpha^{(2)}_{2{\bf q}2\omega;{\bf q}\omega,{\bf q}\omega}$ is an odd function of the chemical potential and {\em does not depend on the electron/hole density} at $|\mu|\gtrsim T$. This is the direct consequence of the linear energy dispersion (\ref{energy-lin}) and essentially differs from the case of the conventional 2D electron gas (\ref{alpha2par-final}) for which $\alpha^{(2)}\propto n_s$. The $q$- and $\omega$-dependencies of $\alpha^{(2)}$ in the linear- and parabolic-spectrum cases are the same, $\alpha^{(2)}\propto q^4/\omega^4$. 

Let us compare the second-order polarizability of graphene (\ref{gr-a2}) with that of a conventional 2D electron gas (\ref{alpha2par-final}), e.g. for typical parameters of 2D electrons in a GaAs/AlGaAs quantum well. Assuming that $m=0.067 m_0$ and $n_s=3\times 10^{11}$ cm$^{-2}$ for GaAs ($m_0$ is the mass of free electron) and $|\mu|\gg T$ for graphene we get
\be 
\frac{\alpha^{(2)}_{graphene}}{\alpha^{(2)}_{GaAs}}\approx 10.\label{ratio-alpha}
\ee 

\section{Self-consistent field\label{sec:self}}

Equations (\ref{a1-def}) and (\ref{a2-def}) determine the first- and second order response of the 2D electron gas to the electric field {\em really acting} on the electrons. Consider now the experimentally relevant formulation of the problem when the system responds to the {\em external} field $\phi_{{\bf q}\omega}^{ext}$. Using the self-consistent field concept we solve, first, the linear response and then the second-order response problem.

\subsection{Linear response}

Consider a 2D electron system under the action of an external electric potential $\phi^{ext}({\bf r},z,t)=\phi_{{\bf q}\omega}^{ext}(z)e^{i{\bf q\cdot r}-i\omega t}$. In the first-order in the external field amplitude the resulting 2D charge density will also contain the ${\bf q}\omega$-harmonic $\rho({\bf r},t)=\rho_{{\bf q}\omega}e^{i{\bf q\cdot r}-i\omega t}$. The density fluctuation creates, in its turn, the induced potential $\phi^{ind}({\bf r},z,t)=\phi_{{\bf q}\omega}^{ind}e^{i{\bf q\cdot r}-i\omega t}$ determined by Poisson equation
\be 
\Delta\phi_{ind}({\bf r},z,t)=-4\pi\rho({\bf r},t)\delta(z)\label{Poisson}
\ee
and given by
\be 
\phi_{{\bf q}\omega}^{ind}(z=0)\equiv\phi_{{\bf q}\omega}^{ind} =\frac{2\pi}q\rho_{{\bf q}\omega}.\label{eq1}
\ee
The density $\rho_{{\bf q}\omega}$ here is determined by the response equation (\ref{a1-def}) in which the really acting on the electrons potential should stay in the right-hand side. This is  not the external field but the {\em total} electric field $\phi_{{\bf q}\omega}^{tot}=\phi_{{\bf q}\omega}^{ext}+\phi_{{\bf q}\omega}^{ind}$ produced both by the external charges and by the 2D electrons themselves. Then we get
\be
\phi_{{\bf q}\omega}^{ind}\equiv\phi_{{\bf q}\omega}^{tot}-\phi_{{\bf q}\omega}^{ext}=\frac{2\pi}q\rho_{{\bf q}\omega}\equiv\frac{2\pi}q\alpha^{(1)}_{{\bf q}\omega;{\bf q}\omega}
\phi_{{\bf q}\omega}^{tot}\label{ind-eps}
\ee
and the known linear-response formula
\be 
\phi_{{\bf q}\omega}^{tot}=\frac{\phi_{{\bf q}\omega}^{ext}}{\epsilon({\bf q},\omega)},\label{resp-lin}
\ee
with the dielectric function
\be 
\epsilon({\bf q},\omega)=1-\frac{2\pi}q\alpha^{(1)}_{{\bf q}\omega;{\bf q}\omega}.
\label{lin-epsilon}
\ee
If the wavevector ${\bf q}$ and the frequency $\omega$ of the external wave satisfy the condition 
\be 
\epsilon({\bf q},\omega)=0,\label{eps=0}
\ee
one gets a resonance in Eq. (\ref{resp-lin}). This resonance corresponds to the excitation of the eigen collective modes of the system -- the 2D plasmons. 

Consider the 2D electron gas with the parabolic energy dispersion. Substituting the linear polarizability (\ref{alpha1par-final}) into Eq. (\ref{lin-epsilon}) we get from (\ref{eps=0}) the known spectrum of the 2D plasmon
\be 
\omega^2=\omega_p^2(q)=\frac{2\pi n_se^2}mq
\ee
first obtained in Ref. \cite{Stern67} (we ignore the dielectric constant of the surrounding medium; if the 2D gas is immersed in the insulator with the dielectric constant $\kappa$, $e^2$ here should be replaced by $e^2/\kappa$).  

In the case of graphene, the spectrum of the 2D plasmons follows from Eqs. (\ref{gr-a1}), (\ref{lin-epsilon}) and (\ref{eps=0}):
\be 
\omega^2=\omega_p^2(q)=\frac {e^2g_sg_v T}{2 \hbar^2 }
\left[ F\left(\frac {|\mu|}{2T}\right) 
+F\left(-\frac {|\mu|}{2T}\right) 
\right]q.\label{plasm-gr}
\ee 
In the limit $|\mu|\gg T$ this gives the result
\be 
\omega_p^2(q)= \frac { e^2g_sg_v |\mu|}{2 \hbar^2 }q \label{plasmonMu}
\ee
obtained in Refs. \cite{Hwang07,Wunsch06}. In the opposite case $|\mu|\ll T$ we get 
\be 
\omega_p^2(q)=\frac { e^2g_sg_vT \ln 2}{ \hbar^2 } q. \label{plasmonT}
\ee
The 2D plasmon problem in the regime $\mu=0$ has been considered in Ref. \cite{Vafek06}. The result reported in \cite{Vafek06} differs from the correct formula (\ref{plasmonT}) by a factor of $4\pi$. 

\subsection{Second order self-consistent response}

Let us now consider the second order response to the external potential 
$\phi_{{\bf q}\omega}^{ext}e^{i{\bf q\cdot r}-i\omega t} + c.c. $ 
The induced and total potential will now contain the frequency harmonics $\pm({\bf q}\omega)$ and $\pm 2({\bf q}\omega)$. The self-consistent charge density then reads
\be 
\rho({\bf r},t)=
\alpha^{(1)}_{{\bf q}\omega;{\bf q}\omega}\phi_{{\bf q}\omega}^{tot}e^{i{\bf q\cdot r}-i\omega t} 
+
\alpha^{(1)}_{2{\bf q}2\omega;2{\bf q}2\omega}\phi_{2{\bf q}2\omega}^{tot}e^{i2{\bf q\cdot r}-i2\omega t} 
+
\alpha^{(2)}_{2{\bf q}2\omega;{\bf q}\omega,{\bf q}\omega} \phi_{{\bf q}\omega}^{tot}\phi_{{\bf q}\omega}^{tot} e^{i2{\bf q\cdot r}-i2\omega t} 
+ c.c., \label{rho-harmonics}
\ee
where the first two terms correspond to the linear response to the first and second harmonics and the third term -- to the second-order response to the first (${\bf q}\omega$) harmonic of the total potential. The complex conjugate terms describe the negative $({\bf q}\omega)$ harmonics. The second order response to $\phi_{{\bf q}\omega}^{tot}\phi_{-({\bf q}\omega)}^{tot} $ vanishes. 

The Fourier harmonics of the induced potential follow from Eq. (\ref{rho-harmonics}) and  Poisson equation (\ref{Poisson}):
\ba 
\phi_{ind}({\bf r},t)&=&
\frac{2\pi}q\alpha^{(1)}_{{\bf q}\omega;{\bf q}\omega}\phi_{{\bf q}\omega}^{tot}e^{i{\bf q\cdot r}-i\omega t}  +
\frac{2\pi}{2q}\alpha^{(1)}_{2{\bf q}2\omega;2{\bf q}2\omega}\phi_{2{\bf q}2\omega}^{tot}e^{i2{\bf q\cdot r}-i2\omega t} + 
\frac{2\pi}{2q}\alpha^{(2)}_{2{\bf q}2\omega;{\bf q}\omega,{\bf q}\omega}\phi_{{\bf q}\omega}^{tot}\phi_{{\bf q}\omega}^{tot} e^{i2{\bf q\cdot r}-i2\omega t} + c.c. 
\nonumber \\ &\equiv &
\phi_{{\bf q}\omega}^{ind}e^{i{\bf q\cdot r}-i\omega t}  +
\phi_{2{\bf q}2\omega}^{ind}e^{i2{\bf q\cdot r}-i2\omega t} + c.c. 
\ea
Equating now the amplitudes of the first-order harmonic (${\bf q}\omega$) we get from here Eqs. (\ref{ind-eps}) and (\ref{resp-lin}). Equating the coefficients at the $(2{\bf q},2\omega)$-harmonic and taking into account that the second harmonic component is absent in the external potential, $\phi_{2{\bf q}2\omega}^{ext}=0$, we get 
\be 
\phi_{2{\bf q}2\omega}^{ind}\equiv \phi_{2{\bf q}2\omega}^{tot}=
\frac{2\pi}{2q}\alpha^{(1)}_{2{\bf q}2\omega;2{\bf q}2\omega}\phi_{2{\bf q}2\omega}^{tot} + 
\frac{\pi}{q}\alpha^{(2)}_{2{\bf q}2\omega;{\bf q}\omega,{\bf q}\omega}\phi_{{\bf q}\omega}^{tot}\phi_{{\bf q}\omega}^{tot}\label{phi-2}
\ee
and finally
\be 
\phi_{2{\bf q}2\omega}^{tot}=
\frac{\pi}{q} 
\frac {\alpha^{(2)}_{2{\bf q}2\omega;{\bf q}\omega,{\bf q}\omega}}
{\epsilon(2{\bf q},2\omega)} \phi_{{\bf q}\omega}^{tot} \phi_{{\bf q}\omega}^{tot} =\frac{\pi}{q} 
\frac {\alpha^{(2)}_{2{\bf q}2\omega;{\bf q}\omega,{\bf q}\omega}}
{\epsilon(2{\bf q},2\omega)\left[\epsilon({\bf q},\omega)\right]^2} \phi_{{\bf q}\omega}^{ext} \phi_{{\bf q}\omega}^{ext}. \label{2nd-res}
\ee

The formula (\ref{2nd-res}), together with (\ref{alpha2par-final}) and (\ref{gr-a2}), represents the main result of this work. One sees that the amplitude of the second-harmonic potential is resonantly enhanced at the frequency $\omega=\omega_p(q)$ [the second-order pole corresponding to the zero of $\epsilon({\bf q},\omega)$] and at the frequency $\omega=\omega_p(q)/\sqrt{2}$ [the first-order pole corresponding to the zero of $\epsilon(2{\bf q},2\omega)$]. These resonances allows one to get a huge enhancement of the second harmonic radiation intensity.

\subsection{Estimates of the second harmonic radiation intensity}

Let us estimate the intensity of the second harmonic signal. Assume that the external potential is
\be 
\phi^{ext}({\bf  r},t)=\phi_0\cos({\bf q\cdot r}-\omega t),\label{phiext}
\ee 
so that $\phi_{{\bf q}\omega}^{ext}=\phi_0/2$. 
Then the total potential at the frequency $2\omega$ obtained from Eq. (\ref{2nd-res}) reads
\be 
\phi^{tot}_{2{\bf q}2\omega}({\bf  r},t)=
\frac{\pi\phi_0^2}{2q} 
\alpha^{(2)}_{2{\bf q}2\omega;{\bf q}\omega,{\bf q}\omega}
\frac {\omega^6 \cos \Big(2({\bf q\cdot r}-\omega t)\Big) }
{\sqrt{\left(\omega^2-\frac{\omega_p^2(2q)}{4}\right)^2+\frac{\omega^2 \gamma^2}4}
\left[\left(\omega^2-\omega_p^2(q)\right)^2+\omega^2\gamma^2\right]},\label{phitot}
\ee 
where we have introduced the momentum scattering rate $\gamma$ in the dielectric function
\be 
\epsilon ({\bf q},\omega)= 1-\frac{\omega_p^2(q)}{\omega(\omega+i\gamma)}
\ee
to remove unphysical divergencies of the plasma resonances. 
Introducing now the intensity of the incident $I_{{\bf q}\omega}^{ext}\sim cE_0^2/8\pi$ and of the second-harmonic wave $I_{2{\bf q}2\omega}^{tot}\sim cE_{tot}^2/8\pi$, where $E_0$ and $E_{tot}$ are the fields corresponding to the potentials (\ref{phiext}) and (\ref{phitot}) respectively, we get the following results:
\begin{enumerate}
\item In the 2D electron gas with the parabolic electron energy dispersion (conventional semiconductor structures)
\be
I_{tot}^{semic} = 
\frac {18 \pi^3 n_s^2e^6 q^4 }{m^4 c}
\frac {\omega^4 }
{\left[\left(\omega^2-\frac{\omega_p^2(q)}{2}\right)^2+\frac{\omega^2 \gamma^2} 4 \right]
\left[\left(\omega^2-\omega_p^2(q)\right)^2+\omega^2\gamma^2\right]^2} I^2_{ext} .\label{semic-intens}
\ee 
\item In the 2D electron gas with the linear energy dispersion (graphene)
\ba 
I_{tot}^{graphene} &=&
\frac {9\pi e^6  v_F^4 q^4}{8 \hbar^4 c} \tanh^2\left( \frac{\mu}{2T} \right)
\frac {\omega^4 
 }
{\left[\left(\omega^2-\frac{\omega_p^2(q)}{2}\right)^2+\frac{\omega^2 \gamma^2}4\right]
\left[\left(\omega^2-\omega_p^2(q)\right)^2+\omega^2\gamma^2\right]^2} I^2_{ext} .\label{graph-intens}
\ea 
\end{enumerate}

\begin{figure}
\includegraphics[width=12cm]{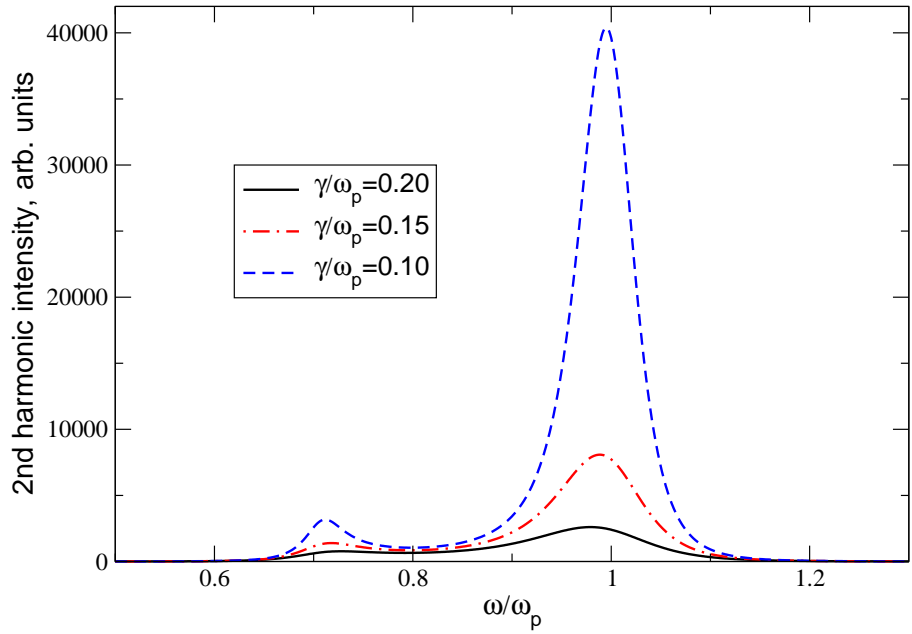}
\caption{\label{fig:omega}The intensity of the second harmonic radiation in semiconductor structures and in graphene as a function of the frequency $\omega/\omega_p(q)$. The parameter of the curves is $\gamma/\omega_p(q)$. }
\end{figure}

\begin{figure}
\includegraphics[width=12cm]{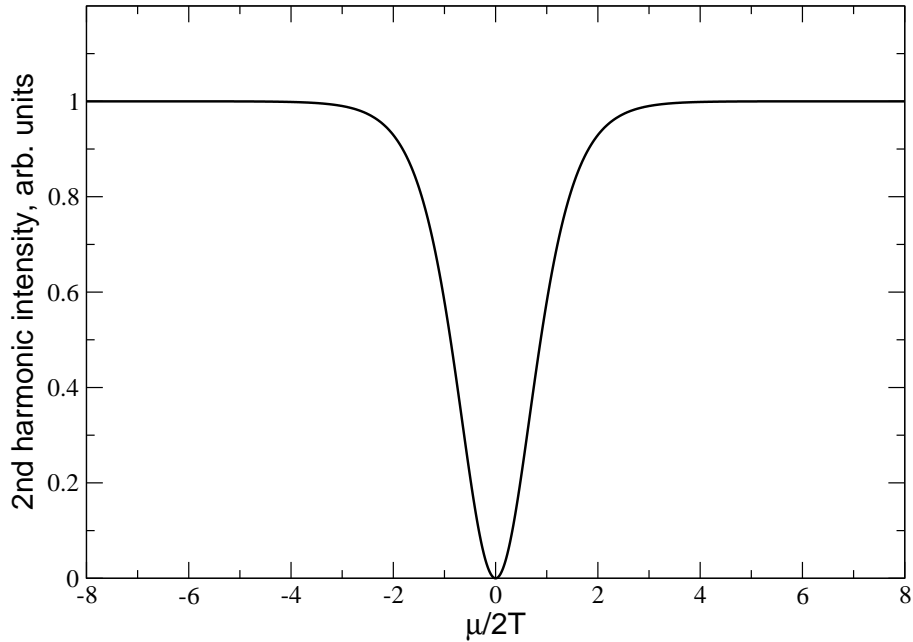}
\caption{\label{fig:mu}The intensity of the second harmonic radiation in graphene as a function of the chemical potential and temperature $\mu/2T$.
}
\end{figure}

The ratio of the intensities (\ref{graph-intens}) and (\ref{semic-intens}) is proportional to the squared ratio of the polarizabilities $\alpha^{(2)}$. For the same parameters that have been used in Eq. (\ref{ratio-alpha}) one gets
\be 
\frac{I_{tot}^{graphene}}{I_{tot}^{GaAs}}\simeq \left(\frac{\alpha^{(2)}_{graphene}}{\alpha^{(2)}_{GaAs}}\right)^2\simeq 100.
\ee
The frequency dependence is the same in both graphene and semiconductor cases and is shown in Figure \ref{fig:omega}. The intensity versus frequency curve has a huge resonance at the frequency $\omega\simeq \omega_p(q)$ and a weaker one  at the frequency $\omega\simeq \omega_p(q)/\sqrt{2}$. The chemical potential and temperature dependence of $I_{tot}^{graphene}$ is shown in Figure \ref{fig:mu}. 

If $\omega=\omega_p(q)$ (the main resonance maximum) the ratio $I_{tot}^{graphene}/I_{ext}$  can be presented as (at $|\mu|\gg T$)
\be 
\frac {I_{tot}^{graphene}}{I_{ext}}
\simeq 
\frac {9\pi e^6  v_F^4 q^4I_{ext}  }{2 \hbar^4 c\omega_p^8(q) } 
\times 
\frac {  \omega_p^4(q) } {\gamma^4} 
\simeq 
\left( \frac {3    E_0   } {16  e k_F^2 } \right)^2
\left(\frac {  \omega_p(q) } {\gamma}\right)^4. 
\ee
In this formula $E_0$ is the electric field of the external incident electromagnetic wave and $e k_F^2$ is the internal electric field in the 2D system (the field produced by an electron at the average inter-electron distance $k_F^{-1}$). The ratio $E_0/ e k_F^2$ in the first brackets is therefore typically very small, $E_0/ e k_F^2\ll 1$. The second factor $\omega_p(q) /\gamma$ is the quality factor of the 2D plasmon resonance, which can be very large in the high-quality samples. This may, at least partly, compensate the smallness of the first factor and substantially facilitate the observation of the second harmonic generation.

\section{Summary\label{sec:summ}}

We have presented the self-consistent analytical theory of the second harmonic generation in two-dimensional electron systems. The theory is applicable to semiconductor structures with the parabolic and graphene with the linear electron energy dispersion. We have shown that the intensity of the second harmonic is about two orders of magnitude larger in graphene than in typical semiconductor structures. Under the conditions of the 2D plasmon resonance the intensity of the second harmonic can be enhanced by several orders of magnitude. 

The frequency of the 2D plasmons in graphene lies in the terahertz range \cite{Liu08,Jablan09,Langer10}. The discussed phenomena can be used for creation of novel devices (frequency multipliers, mixers, lasers) operating in this technologically important part of the electromagnetic wave spectrum.

\acknowledgments

The financial support of this work by Deutsche Forschungsgemeinschaft is gratefully acknowledged.

%\bibliography{../../BIB-FILES/zerores,../../BIB-FILES/emp,../../BIB-FILES/lowD,../../BIB-FILES/mikhailov,../../BIB-FILES/math,../../BIB-FILES/graphene}
%\bibliographystyle{apsrev}

\end{document}